\documentclass[preprint,aps,prd,nofootinbib,showpacs,superscriptaddress,11pt]{revtex4}
\usepackage{txfonts}
\usepackage[colorlinks, linkcolor=blue, citecolor=blue, urlcolor=blue]{hyperref}
\usepackage{float}
\usepackage{graphicx}
\usepackage{dcolumn}
\usepackage{bm}
\usepackage{amssymb}
\usepackage{latexsym}

\begin{document}

\title{Cosmological constraints on the new holographic dark energy model with action principle}

\author{Miao Li}
\email{mli@itp.ac.cn}
\affiliation{Institute of Theoretical Physics, Chinese Academy of Sciences, Beijing 100190, China}
\affiliation{Kavli Institute for Theoretical Physics China, Chinese Academy of Sciences, Beijing 100190, China}
\affiliation{State Key Laboratory of Frontiers in Theoretical Physics, Chinese Academy of Sciences, Beijing 100190, China}

\author{Xiao-Dong Li}
\email{renzhe@mail.ustc.edu.cn}
\affiliation{Institute of Theoretical Physics, Chinese Academy of Sciences, Beijing 100190, China}
\affiliation{Korea Institute for Advanced Study, Hoegiro 87, Dongdaemun-Gu, Seoul 130 - 722, Republic of Korea}

\author{Jun Meng}
\email{mengjun@itp.ac.cn}
\affiliation{Institute of Theoretical Physics, Chinese Academy of Sciences, Beijing 100190, China}
\affiliation{Kavli Institute for Theoretical Physics China, Chinese Academy of Sciences, Beijing 100190, China}
\affiliation{State Key Laboratory of Frontiers in Theoretical Physics, Chinese Academy of Sciences, Beijing 100190, China}

\author{Zhenhui Zhang}
\email{zhangzhh@mail.ustc.edu.cn}
\affiliation{Institute of Theoretical Physics, Chinese Academy of Sciences, Beijing 100190, China}

\begin{abstract}
Recently, a New HDE model with action principle was proposed \cite{NewHDE}.
It is the first time that the holographic dark energy model is derived from the action principle.
This model completely solves the causality and circular problems in the original HDE model,
and automatically gives rise to a dark radiation component.
Thus, it is worth investigating such an interesting model by confronting it with the current cosmological observations,
so that we can check whether the model is consistent with the data,
and determine the regions of parameter space allowed.
These issues are explored in this work.
Firstly, we investigate the dynamical behaviors and the cosmic expansion history of the model,
and discuss how they are related with the model parameter $c$.
Then, we fit the model to a combination of the present Union2.1+BAO+CMB+$H_0$ data.
We find the model yields  $\chi^2_{\rm min}=548.798$ (in a non-flat Universe),
comparable to the results of the original HDE model (549.461) and the concordant $\Lambda$CDM model (550.354).
At 95.4\% CL, we get $1.41<c<3.09$ and correspondingly $-2.25<w(z=-1)<-1.39$,
implying the Big Rip fate of the Universe at a high confidence level.
Besides, for the constraints on dark radiation, 
we also get a rough estimation $N_{\rm \rm eff}=3.54^{+0.32+0.67}_{\rm -0.45-0.76}$,
with the central value slightly larger than the standard value 3.046.
\end{abstract}


\maketitle

\section{Introduction}

Since the discovery of cosmic acceleration \cite{Riess},
dark energy has become one of the most popular research areas in cosmology \cite{DEReview}.
Numerous dark energy models have been proposed in the last decade.
However, the nature of dark energy still remains a mystery.

Actually, the dark energy problem may be in essence an issue of quantum gravity \cite{Witten:2000zk}.
In the absent of a complete theory of quantum gravity,
the most plausible approach is to consider some effective theories,
in which some fundamental principles are taken into account.
It is commonly believed that the holographic principle is a fundamental principle of quantum gravity \cite{Holography}.
In \cite{Cohen}, based on the effective quantum field theory,
Cohen {\it et al.} suggested that quantum zero-energy energy of a system with size $L$ shall not exceed the mass of a black hole with the same size, i.e.,
\begin{equation}
 L^3 \Lambda^4\leq L M^2_{p},
\end{equation}
here $\Lambda$ is the ultraviolet (UV) cutoff which is closely related to the zero-point energy density,
and $M_p\equiv 1/\sqrt{8\pi G}$ is the reduced Planck mass.
In this way, the UV cutoff of the system is related to its infrared (IR) cutoff.
When we take the whole universe into account, 
the vacuum energy related to this holographic principle can be viewed as dark energy.
The largest IR cutoff $L$ is chosen by saturating the inequality, so we get the dark energy density
\begin{equation}
 \rho_{de}=3c^2M^2_pL^{-2}
\end{equation}
where $c$ is a dimensionless model parameter.
In \cite{Li1}, Li suggested to choose the future event horizon of the universe as the IR cutoff of this theory, defined as,
\begin{equation}\label{eq:Rh}
 R_h=a\int_t^{+\infty}\frac{dt}{a}.
\end{equation}
This choice not only gives a reasonable value for the dark energy density,
but also leads to an accelerated universe.

The holographic dark energy (HDE) model based on Eq. (\ref{eq:Rh}) has been proved to be a promising dark energy candidate. 
The the original paper \cite{Li1}, Li showed that the HDE model can explain the coincidence problem.
In \cite{HDEstable}, it is proved that the model is perturbation stable.
Following studies also show that the model is in good agreement with the current cosmological observations \cite{HDEObserv}.
Thus, the HDE model becomes one of the most competitive and popular dark energy candidates,
and attracts a lot of interests \cite{HDEworks}.

In spite of its success, 
the HDE model still suffers from some criticisms \cite{Kim},
due to its use of the future event horizon as the present cutoff.
In this model, the evolution of the universe depends on the future information about the universe,
so there is a causality problem.
And, the future event horizon exists only in an accelerating universe,
so there is a circular logic problem if one use 
an assumption based on the accelerating expansion to explain the accelerating expansion.
Besides, the lack of derivation of the model from an action is also a blemish.
These problems remain unsolved since the proposal of the model in 2002.

Fortunately, these problems are solved in a recent work of Li and Miao \cite{NewHDE}.
For the first time, they derived a holographic dark energy model
(hereafter the New HDE model) from the action principle.
In the new model, the evolution of the universe only depends on the present state of the universe, 
clearly showing that it obeys the law of causality. 
Furthermore, in the new model the use of future event horizon as a present cut-off is not an input but automatically follows from equations of motion.
So the puzzles of causality and circular logic are all completely solved. 
In \cite{NewHDE}, the authors showed that the New HDE model is very similar to the original HDE model,
except a new term which may be interestingly explained as dark radiation \cite{darkradiation}.

It is worth investigating such an interesting model by confronting it with the current cosmological observations.
That will enable us to answer a lot of interesting questions:
Is the model consistent with the current data?
What regions of parameters space are allowed by data?
In the original HDE model, the big-rip fate of the universe is favored by the data, 
what is the fate of universe in the New HDE models?
and so on.
These issues are not covered in \cite{NewHDE}, and will be explored in this work.

This paper is organized as follows.
In Sec. II, we give a brief introduction to the New HDE model.
In Sec. III, we investigate the dynamical behaviors and the cosmic expansion history of the model,
and discuss how they are related with the model parameter $c$.
In Sec. IV, we fit the model to the combined Union2.1+BAO+CMB+$H_0$ data,
and present the fittings results.
Many interesting issues, including the EoS of the New HDE, 
the fate of the universe, the dark radiation, are discussed.
Some concluding remarks are given in Sec. V. 
In this work, we assume today's scale factor $a_0 = 1$, 
so the redshift $z$ satisfies $z = 1/a - 1$.
We use negative redshift to denote the future,
and $z=-1$ corresponds the far future $a=\infty$.
The subscript ``0'' indicates the present value of the corresponding quantity unless otherwise specified.

\section{A Brief Introduction to the New HDE Model}%modified

In this section, we briefly introduce the New HDE model proposed in \cite{NewHDE}
\footnote{We only focus on the model with the future event horizon as cut-off,
which can lead to cosmic acceleration}.

\subsection{Derivation of the Model from the Action Principle}

Following \cite{NewHDE}, we review how the model is derived from the action principle.
Consider the Robertson-Walker metric
\begin{eqnarray}\label{metric}
ds^2=-N^2(t)dt^2+a^2(t)[\frac{dr^2}{1-k r^2}+r^2d\Omega^2]
\end{eqnarray}
and the action
\begin{eqnarray}\label{action}
S=\frac{1}{16\pi G}\int dt[\sqrt{-g}(R-\frac{2c}{a^2(t)
L^2(t)})-\lambda(t)(\dot{L}(t)+\frac{N(t)}{a(t)})]+S_{\rm M},
\end{eqnarray}
where $R$ is the Ricci scalar, $k$ represents for curvature,
$\sqrt{-g}=N a^3$ (we have integrated the $r, \theta, \phi$ parts),
and M denotes the action of all matter fields
(we use m to denote the matter without pressure and r to denote the radiation).
By taking the variations of $N, a, \lambda, L$, and redefining $Ndt$ as $dt$, 
we obtain
\begin{eqnarray}\label{equation0}
(\frac{\dot{a}}{a})^2+\frac{k}{a^2}=\frac{c}{3a^2L^2}+\frac{\lambda}{6a^4}+\frac{8\pi}{3}\rho_{\rm M},\nonumber \\
\frac{2\ddot{a}a+\dot{a}^2+k}{a^2}=\frac{c}{3a^2L^2}-\frac{\lambda}{6a^4}-8\pi p_{\rm M},
\end{eqnarray}
and
\begin{eqnarray}\label{equation1}
\dot{L}&=&-\frac{1}{a},\ \ \ \ \ L=\int_t^{\infty}\frac{dt'}{a(t')}+L(a=\infty)\nonumber, \\
\dot{\lambda}&=&-\frac{4ac}{L^3},\ \
\lambda=-\int_0^{t}dt'\frac{4a(t')c}{L^3(t')}+\lambda(a=0).
\end{eqnarray}
It follows the holographic dark energy density
\begin{equation}\label{Eq:rhohde}
\rho_{\rm hde}=\frac{1}{8\pi G}\left(\frac{c}{a^2L^2}+\frac{\lambda}{2a^4}\right),
\end{equation}
which is characterized by the horizon $aL$, and a new term $\frac{\lambda}{2a^4}$.
In this term, the $\lambda(a=0)$ component evolves in the same way as radiation,
thus can be naturally interpreted as dark radiation \cite{darkradiation}.

In \cite{NewHDE}, by deriving the asymptotic solutions of the equations,
the authors proved that 
\begin{equation}
 L(a=\infty)=0,
\end{equation}
so $aL$ is exactly the future event horizon.

Before going on, we mention that the functions $L(t)$, $\lambda(t)$, and $N(t)$ all have clear physical meanings.
Their values are related with observable quantities,
and can not be arbitrarily rescaled:
$aL$ is the size of the future event horizon.
From Eq. (\ref{equation1}), both $L$ and $\lambda$ can be determined through measurements of the cosmic expansion history.
The $\lambda(a=0)$ term behaves the same as radiation,
thus can be naturally interpreted as dark radiation \cite{darkradiation}.
We can determine its value by measuring the amount of dark radiation.
$N(t)dt$ determines the time component of the metric. 
Since we require that $t$ is the comoving time, it follows that $N(t)=1$.

Let us give some comments on the New HDE model:
First, it is the first holographic dark energy derived from action principle.
Second, as shown in Eqs. (\ref{equation0},\ref{equation1}),
in this model the evolution of the universe only depends on the present initial conditions
$a,\ \dot a,\ L,\ \lambda$, so there is no causality problem.
Third, in this model, the use of future event horizon as the cut-off is not an input. 
Instead, it follows automatically from the equations of motion.
So the logical circular problem is also solved.

Finally, as preparations for the numerical analysis,
let us rewrite the equations in the redshift space.
Let us define the ``Hubble-free'' quantities
\begin{eqnarray}
\tilde{L}\equiv H_0L,\ \ \tilde{\lambda}\equiv\lambda/H_0^2,\ \ E(z)\equiv\frac{H}{H_0},
\end{eqnarray}
utilizing the fact that $\frac{d}{dt}=-H(1)z)\frac{d}{dz}$,
we rewrite Eq. (\ref{equation1}) as
\begin{equation}\label{Eq:LlambdaEqs}
\frac{d\tilde{L}}{dz}=\frac{1}{E(z)},\ \
\frac{d\tilde{\lambda}}{dz}=\frac{4c}{(1+z)^2E(z)\tilde{L}^3}.
\end{equation}
From Eq. (\ref{equation0}), $E(z)$ takes the form
\begin{equation}\label{Eq:Ez}
E(z)=\sqrt{\Omega_{\rm m0}(1+z)^3+\Omega_{\rm k0}(1+z)^2+\Omega_{\rm r0}(1+z)^4
+\frac{1}{3}\left(\frac{c(1+z)^2}{\tilde{L}^2}+\frac{\tilde{\lambda}(1+z)^4}{2}\right)},
\end{equation}
where $\Omega_{\rm m0}$, $\Omega_{\rm k0}$, $\Omega_{\rm r0}$ are the current ratios of matter, curvature, radiation.
Here $\Omega_{\rm r0}$ includes the components of photon and neutrino, given by
\begin{equation}\label{Eq:omegar}
\Omega_{\rm r0}=\Omega_{\rm \gamma 0}(1+0.2271N_{\rm \rm eff,sd}),\ \Omega_{\rm \gamma 0}=2.469 \times 10^{-5} h^{-2},
\end{equation}
where $\gamma$ represents for photons,
and $N_{\rm \rm eff,sd}=3.046$ is the ``standard'' value of effective number of neutrino species \cite{Neffsd}.
The last term in the square-root of Eq. (\ref{Eq:Ez}) include both the dark energy and dark radiation components,
with the ratio
\begin{equation}
\Omega_{\rm de+dr}(z)=\frac{1}{3E(z)^2}\left(\frac{c(1+z)^2}{\tilde{L}^2}+\frac{\tilde{\lambda}(1+z)^4}{2}\right).
\end{equation}

\subsection{Dark Energy Equation of State}\label{subsec:DEEoS}

The EoS of the New HDE takes the form \cite{NewHDE}
\begin{equation}
w\equiv\frac{p_{\rm hde}}{\rho_{\rm hde}}=\frac{\lambda L^2-2ca^2}{3\lambda L^2+6ca^2}=\frac{\tilde{\lambda} \tilde{L}^2-2ca^2}
{3\tilde{\lambda} \tilde{L}^2+6ca^2}.
\end{equation}
Similar to the original HDE model,
its property is closely related with the value of $c$.
Since the Big Rip problem is a hot topic when people investigated the original HDE 
(see, e.g., Ref. \cite{HealWorld} and references therein),
let us have a look at the asymptotic behavior of the New HDE EoS when $z\rightarrow-1$ \cite{NewHDE}
\begin{equation}\label{Eq:w-1}
w(z=-1)=\frac{-3+2c+\sqrt{9+12c}}{3(-3-2c+\sqrt{9+12c})}.
\end{equation}
\begin{figure}[H]
\centering{
\includegraphics[height=6cm]{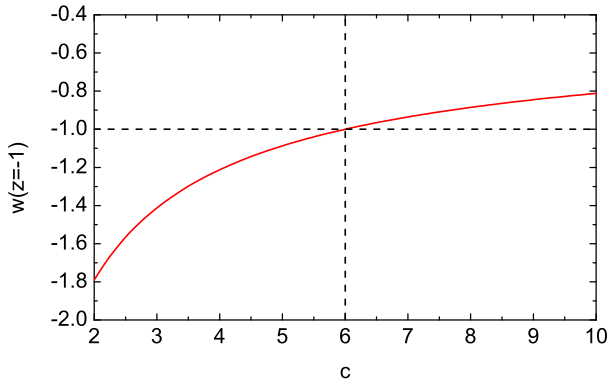}}
\caption{\label{Fig:w(-1)}
The dependence of $w(z=-1)$ on the parameter $c$.
$w=-1$, $c=6$ are plotted in black dashed lines.}
\end{figure}
We plot this relation in Fig. \ref{Fig:w(-1)}.
For $c<6$, we have $w(z=-1) < -1$,
and the Universe will end up with a Big Rip.
For $c>6$, we have $w(z=-1) > -1$, and the future behavior of New HDE is quintessence like.
The asymptotical de Sitter solution is accomplished when $c=6$.

At high redshift, if we neglect the $\lambda(a=0)$ term (the dark radiation component) in Eq. (\ref{equation1})
and focus on the dark energy component proportional to $(aL)^{-2}$,
asymptotically we have
\begin{equation}
w\rightarrow-\frac{1}{3}
\end{equation}
which is also similar to the original HDE model.

\subsection{Dark Radiation}\label{subsec:DR}

The $\lambda(a=0)$ term in Eq. (\ref{equation1}) corresponds to an energy component satisfying
$\rho\propto a^{-4}$ and $p=\frac{1}{3}\rho$,
thus can be naturally explained as dark radiation.

Notice that in our convention  $\Omega_{\rm r0}$ in Eq. (\ref{Eq:Ez}) only covers photon and neutrino.
To describe the dark radiation component, let us define
\begin{equation}
\rho_{\rm r+dr}\equiv\rho_{\rm r}+\rho_{\rm dr}\equiv\rho_{\rm \gamma}(1+0.2271N_{\rm \rm eff})
\end{equation}
and characterize the dark radiation (labeled as ``dr'') by the parameter $N_{\rm \rm eff}$.
It follows straightforward that the dr-r ratio is
\begin{equation}
\frac{\rho_{\rm dr}}{\rho_{\rm r}}=\frac{0.2271(N_{\rm \rm eff}-N_{\rm \rm eff,sd})}{1+0.2271N_{\rm \rm eff,sd}},
\end{equation}
where $N_{\rm eff,sd}=3.046$ as mentioned above.
On the other hand, from Eq. (\ref{Eq:rhohde}),
in the radiation dominate epoch (denoted by $z_{\rm rd}$) we have
\begin{equation}\label{Eq:drrratio}
\frac{\rho_{\rm dr}}{\rho_r} \approx \frac{\tilde{\lambda}(z_{\rm rd})}{6\Omega_{\rm r0}}.
\end{equation}
Thus, from the above two equations, $N_{\rm \rm eff}$ is determined by
\begin{equation}\label{Eq:Neff}
N_{\rm \rm eff}=N_{\rm \rm eff,sd}+\frac{1+0.2271N_{\rm \rm eff,sd}}{0.2271}\left(\frac{\tilde{\lambda}(z_{\rm rd})}{6\Omega_{\rm r0}}\right).
\end{equation}
Using the fact that $\Omega_{\rm r0}\approx10^{-4}$ and $N_{\rm \rm eff}\approx3-5$ (see e.g. \cite{darkradiation}),
roughly we require
\begin{equation}\label{Eq:lambdacon}
|\tilde{\lambda}(z_{\rm rd})|\lesssim10^{-4}
\end{equation}
to be consistent with the cosmological observations.

\subsection{The Set of Free Parameters}

Here we list the full set of free parameters of the New HDE model is
\footnote{$\Omega_{\rm r0}$ is determined by $h$ through Eq. (\ref{Eq:omegar}), so it is a derived parameter.
Notice that $N_{\rm eff}$ is also a derived parameter:
Given the values of the five parameters in Eq. (\ref{Eq:ParSet}),
we can solve the derivative equations Eq. (\ref{Eq:LlambdaEqs}), obtain the value of $\tilde{\lambda}(z_{\rm rd})$,
and use Eq. (\ref{Eq:Neff}) to obtain the value of $N_{\rm eff}$.
To make our fitting complete and reliable, we do not assume a flat background and include $\Omega_{k0}$ into the set of parameters 
(see \cite{Bassett} for reference).},
\begin{equation}\label{Eq:ParSet}
 {\bf P}=\{\Omega_{\rm m0}, \Omega_{\rm k0}, c, \tilde{L}_0, h\}.
\end{equation}
Compared with the original HDE model without dark radiation component \cite{Li1},
the New HDE model has one extra parameter $\tilde{L}_0$, 
the current size of the reduced future event horizon.
Notice that the New HDE model considered in our analysis has 5 model parameters,
which is 3 more than the minimal stadard $\Lambda$CDM model (in a flat Universe, with standard number of neutrino species).

In this work, we numerically solve Eq.(\ref{Eq:LlambdaEqs}) to obtain background evolutions of the New HDE model
\footnote{Although the equations of motion have exact analytical solutions $a(L)$ \cite{NewHDE},
technically it is much easier to obtain $L(z)$ and $\lambda(z)$ by numerically solving the derivative equations \cite{Wolfram}.}.
To complete the equations, two initial conditions are required.
One initial condition comes from the fact that the high-redshift value of $\tilde{\lambda}(z)$ is determined by the dark radiation component,
\begin{equation}\label{Eq:lambdazrd}
\tilde{\lambda}(z_{\rm rd})=6\Omega_{\rm r0}(N_{\rm eff}-N_{\rm eff,sd})\frac{0.2271}{1+0.2271N_{\rm eff,sd}},
\end{equation}
and the other initial condition follows from the relation $E(z=0)=1$,
\begin{equation}\label{Eq:inicon}
\frac{\tilde{\lambda}_0}{2}+\frac{c}{\tilde{L}_0^2}=3(1-\Omega_{\rm m0}-\Omega_{\rm k0}-\Omega_{\rm r0}).
\end{equation}

\section{Dynamical Behaviors and the Cosmic Expansion History}%modified

In this section we discuss the dynamical behaviors and the predicted cosmic expansion histories of the New HDE model,%modified
divided into the $c<6$ and $c\geq 6$ cases.

\subsection{Dynamical Behaviors}\label{Sec:DynamicalBehaviors}

Representatively, let us take six values of $c=2.3,\ 2.4,\ 2.5,\ 6,\ 7,\ 8$,
and investigate the dynamical properties of the model in $c<6$ and $c\geq 6$ cases.

%\footnote{
%This set of parameters corresponds to
%$\Omega_{\rm r0}=7.42564\time10^{-5}$,
%$\Omega_{\rm de0}=0.749926$,
%and $\tilde{\lambda}_0=-3.678/-4.034/-4.389$ for $c=2.3/2.4/2.5$.}.

\begin{figure}[H]
\centering{
\includegraphics[width=16cm]{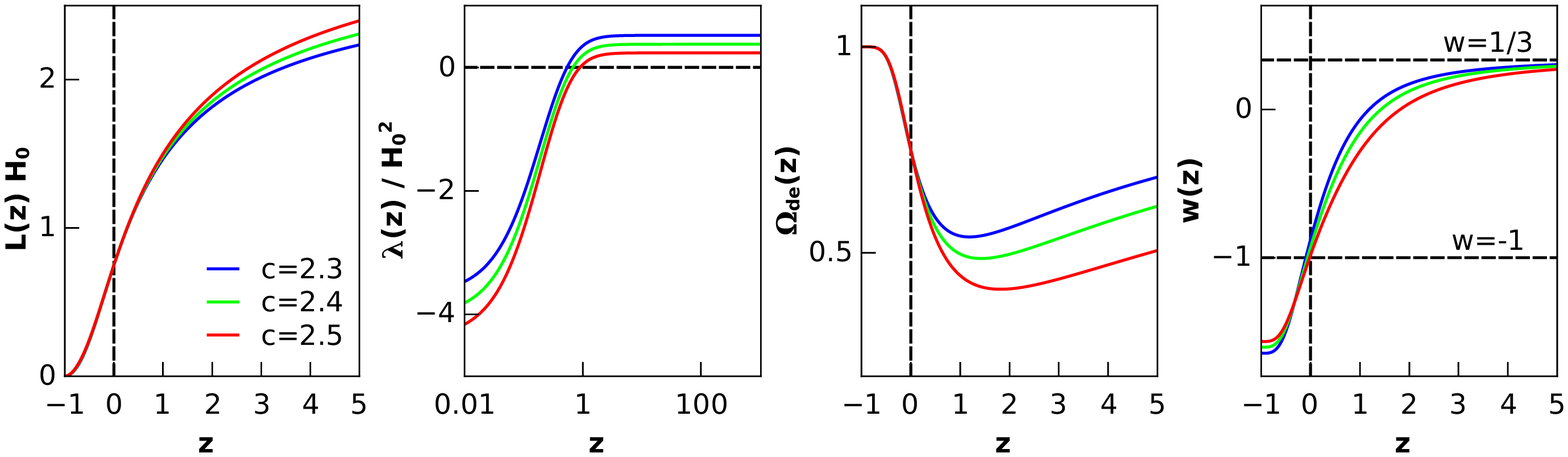}
\includegraphics[width=16cm]{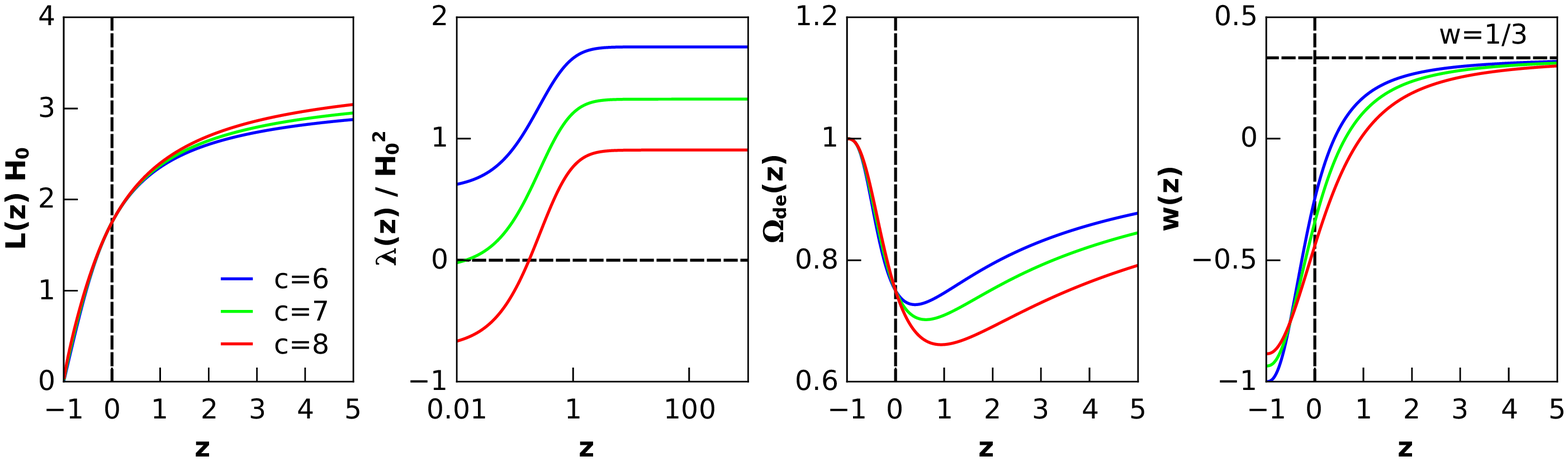}
}
\caption{\label{Fig:Llambdaz}
Evolutions of $\tilde{L}(z)$, $\tilde{\lambda}(z)$, $\Omega_{\rm de+dr}(z)$ and $w(z)$.
We choose the current size of future event horizon $\tilde{L}_0$=0.75 for the $c<6$ models, 
and $\tilde{L}_0$=1.75 for the $c>6$ models.
{\it Upper panels}: The $c$=2.3 (blue),\ 2.4 (green),\ 2.5 (red) cases.
{\it Lower panels}: The $c$=6 (blue),\ 7 (green),\ 8 (red) cases.}
\end{figure}

In Fig. \ref{Fig:Llambdaz}, we plot the evolutions of $\tilde{L}(z)$, $\tilde{\lambda}(z)$, $\Omega_{\rm de+dr}(z)$ and $w(z)$ for the six different cases.
$c<6$ cases are shown in upper panels, while $c\geq6$ cases are shown in lower panels.
The current size of reduced future event horizon is chosen as $\tilde{L}_0=0.75$ for the $c<6$ cases, $\tilde{L}_0=1.75$ for the $c\geq6$ cases.
For the other parameters, we fix $\Omega_{\rm m0}=0.25,\ \Omega_{\rm k0}=0,\ h=0.75$.

Many interesting phenomena are found in this figure.
First, as proven in \cite{NewHDE},
we find that the numerical analysis shows $L(z\rightarrow-1)=0$.
That is, the equations of motion force $aL$ to be exactly the future event horizon.
Second, we find the future behavior of New HDE is phantom like for the $c<6$ cases,
cosmological constant like for the $c=6$ case,
and quitessence like for the two $c>6$ cases.
We find $w(z=-1)=-1.645/-1.604/-1.566/-1/-0.935/-0.885$ for $c=2.3/2.4/2.5/6/7/8$,
in good agreement with the analytical result Eq. (\ref{Eq:w-1}).
Third, in all cases we find $\tilde{\lambda}>0$ at high redshift,
corresponding to a positive dark radiation component
\footnote{
Here the amount of dark radiation is so much that dark radiation becomes dominated even at low redshift
(as seen in the plottings of $\Omega_{\rm de+dr}(z)$ and $w(z)$,
the ratio of New HDE becomes increasing at $z\sim1-2$, and $w(z)\rightarrow\frac{1}{3}$ at $z\sim5$).
The reason is that we failed to choose a proper set of parameters satisfying Eq. (\ref{Eq:lambdacon}).
This problem will certainly be solved through data-fitting in the next section.}.
Finally, we find the present value of dark energy EoS $w_0\sim$-1 for the three $c<$ cases, 
and while for the three $c\geq$6 cases there are $-0.5\lesssim w_0\lesssim-0.25$.
Comparably, the three $c<6$ cases are more consistent with current cosmological observations.

\subsection{The Expansion History}

The aim of this work is to confront the New HDE model with the observational data of the cosmic expansion history.
So, before fitting the model to the data, it is worth investigating the expansion history of the model.

\begin{figure}[H]
\centering
\includegraphics[scale=0.35]{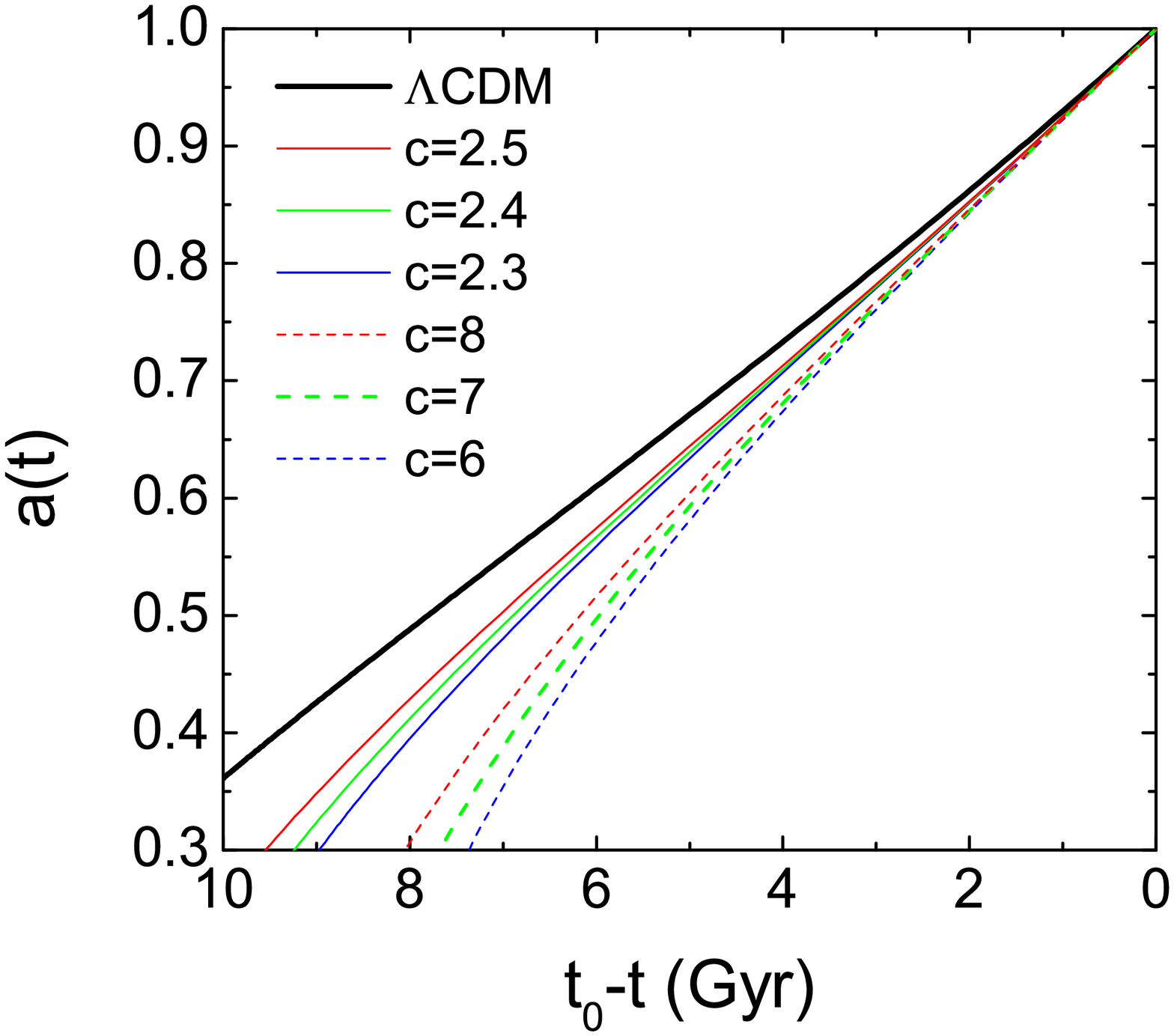}
\includegraphics[scale=0.35]{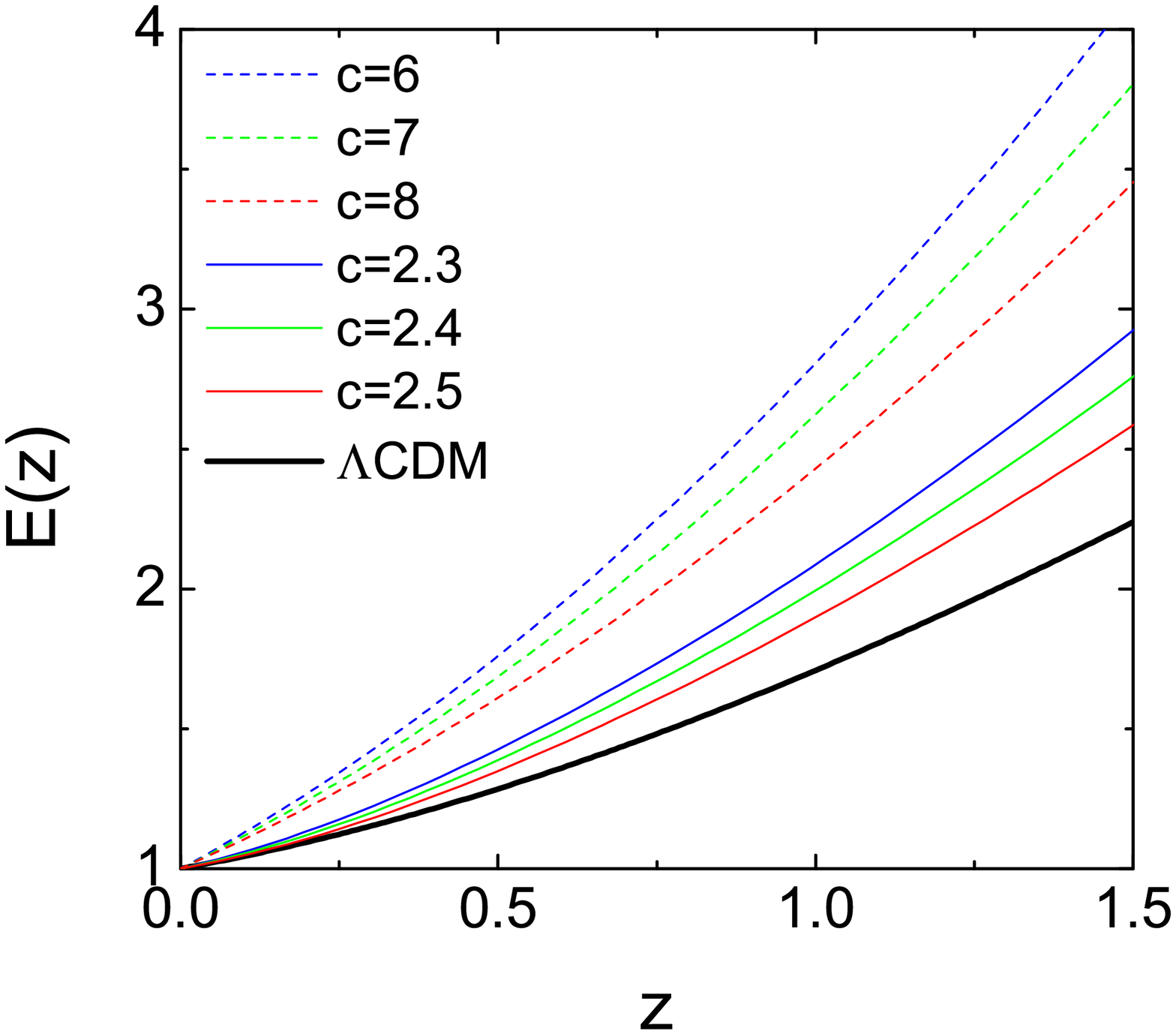}
\caption{\label{Fig:expansionhistory}
The expansion history given by different model parameters.
Solid lines show the $c$=2.3 (blue), $c$=2.4 (green), $c$=2.5 (red) cases,
while the dashed lines show the $c$=6 (blue), $c$=7 (green), $c$=8 (red) cases.
The current size of future event horizon $\tilde{L}_0$ is chosen as 0.75 for the $c<6$ models and 1.75 for the $c\geq6$ models.
For comparison, the $\Lambda$CDM model with WMAP7 best-fit parameters is plotted in thick black line.
{\it Left panel}: The scale factor $a(t)$ as a function of comoving time.
%The horizonal coordinate $t_0-t$ is the interval between the past time and present time.
{\it Right panel}: The reduced Hubble parameter $E(z)$ as a function of redshift.}
\end{figure}

In Fig. \ref{Fig:expansionhistory},
we plot the expansion histories of the models with six sets of parameters discussed in the previous subsections.
The left panel shows the evolution of the scale factor $a(t)$ as a function of time,
while the right panel shows the evolution of the reduced Hubble parameter $E(z)$ as a function of redshift.
For comparison, we also plot the $\Lambda$CDM model with the WMAP7 best-fit parameters (Talbe 1 of \cite{WMAP7}) in thick black line.
The $c<6$ and $c>6$ cases are plotted in solid and dashed lines, respectively.

A most evident phenomenon in the figure is that the three $c>6$ cases (dashed lines) deviate a lot from the $\Lambda$CDM model,
implying that they are inconsistent with the cosmological observations.
This is expectable from Fig. \ref{Fig:Llambdaz}, 
where we see these models have large values of dark energy EoS.
This means a high dark energy density in the past,
and thus a higher expansion rate in the past according to the Friedmann Eq. (\ref{equation0}).
Compared with the $\Lambda$CDM model, these models give larger expansion rate $E(z)$s,
and correspondingly more rapidly increasing $a(t)$s.

Contrastingly, the expansion history of the three $c<6$ models (the solid lines) are more closed to the $\Lambda$CDM model.
Among the six cases considered, the parameter set $c=2.5$, $\tilde{L}_0=0.75$ provides an expansion history most closed to the $\Lambda$CDM model,
although evident discrepancy still exists.

Besides, these plottings show that with fixed $\tilde{L}_0$ smaller $c$s always yield to higher expansion rates.
The reason is that in the $z>0$ region $w$ increases when we decrease the value of $c$, 
as shown in Fig. \ref{Fig:Llambdaz}.

\begin{figure}[H]
\centering
\includegraphics[scale=0.35]{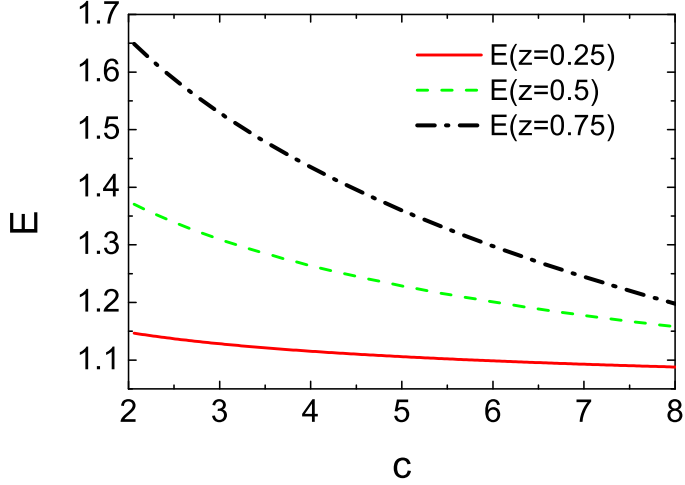}
\caption{\label{Fig:Ec}
The relation between the expansion history and the model parameter $c$.
The values of $E(z)$ at $z=0.25$ (red solid), $z=0.5$ (gren dashed) and $z=0.75$
(black dash dotted) are plotted.
In all plottings, we choose $\Omega_{\rm m0}=0.25,\ \Omega_{\rm k0}=0,\ h=0.75$ and
$c/\tilde{L}^2_0=0.45$.}
\end{figure}

To investigate the effect of $c$ on the expansion history,
in Fig. \ref{Fig:Ec} we show the dependence of $E(z)$ on $c$ at three redshifts $z=0.25,\ 0.5$, and 0.75.
In all plottings, we choose $\Omega_{\rm m0}=0.25,\ \Omega_{\rm k0}=0,\ h=0.75$ and
$c/\tilde{L}^2_0=0.45$
\footnote{
According to Eq. \ref{Eq:inicon}, fixing $c/\tilde{L}^2_0=0.45$ means requiring $\tilde{\lambda}_0\approx-4.5$.
Actually, we find $c/\tilde{L}^2_0=0.45$ roughly describes the shape of the $c$-$\tilde{L}$ parameter space 
constrained by the cosmological data
(see the lower-right panel of Fig. \ref{Fig:ConLike}).
}.
With the above parameters fixed,
we find that $E$ is a monotonic decreasing function of $c$ at a given redshift.
Especially, an $\Omega_m=0.25$ $\Lambda$CDM model
(with $E(z)$=$1.11$, $1.26$, $1.45$ at $z=0.25$, $0.5$, $0.75$)
correponds to the values of $c\approx 3-4$.

\section{Cosmological Constraints from the Obervational Data}

In this section we discuss the cosmological interpretations of the New HDE model
by confronting it with the cosmological observations.
The data used in our analysis include:
\begin{itemize}
\item
The Union2.1 sample of 580 SNIa \cite{Union2.1}.
\item
The ``WMAP distance priors'' given by the 7-yr WAMP observations \cite{WMAP7},
including the ``acoustic scale'' $l_A$ and the ``shift parameter'' $R$.
\item
For the BAO data, we use the measurements of $r_s(z_d)/D_V(0.2)$ and $r_s(z_d)/D_V(0.35)$
from the SDSS DR7 \cite{SDSSDR7},
the measurement of $r_s(z_d)/D_V(0.106)=0.336\pm0.015$ given by 6dFGS \cite{6dFGS}.
We also use the measurements of the acoustic parameter 
$A(z)\equiv \frac{100 D_V(z)\sqrt{\Omega_{m0}h^2}}{z}$ \cite{AcouP}
at $z=0.44,\ 0.6,\ 0.73$ from the WiggleZ Dark Energy Survey \cite{WiggleZ}.
\item
The Hubble constant measurement $H_0=73.8\pm 2.4 {\rm km/s/Mpc}$ from the WFC3 on the HST \cite{HSTWFC3}.
\end{itemize}
Notice that the data set we used are exactly the same as the data used in Ref. \cite{IHDE},
so we can compare the fitting results of the New HDE model with the results of the original HDE model and
the concordant $\Lambda$CDM model obtained in \cite{IHDE}.

We combine the above data to perform the $\chi^2$ analysis.
For simplicity, we will not explain the cosmological data and the $\chi^2$ analysis in detail
(for a detailed description, see Ref. \cite{IHDE}).
Here we only mention that the data can put interesting constraints on the cosmic expansion history
in both low-redshift and high-redshift regions.
On one hand, the SNIa data, the BAO $A$ parameter, the CMB $R$ parameter and the $H_0$ measurement
are powerful at low redshift region,
when the dark energy component is important.
On the other hand, at high redshift region when the dark radiation component is important,
we adopt the measurements of $r_s(z_d)/D_V(z)$ and $r_s(z_*)$ from the BAO and CMB observations,
where the comoving sound horizon $r_s$ takes the form
\begin{equation}
r_s(z)=\frac{1} {\sqrt{3}}  \int_0^{1/(1+z)}  \frac{ da } { a^2H(a)
\sqrt{1+(3\Omega_{\rm b0}/4\Omega_{\rm \gamma0})a} }~.
\label{eq:rs}
\end{equation}
This quantity encodes the information of the Hubble parameter $H(z)$ at high redshift,
thus can lead to valuable constraints on the dark radiation amount.

In the following we will discuss the cosmological constraints on the New HDE model,
divided into three subsections:
We present the fitting results in the first subsection,
and specifically discuss the Equation of State and Dark Radiation in the other two subsections.

\subsection{Fitting Results}

Using the above data, we find the goodness-of-fit of the New HDE model is
\begin{equation}
\chi^2_{\rm min}=548.798.
\end{equation}
This result is comparable to the results of the original HDE model (549.461)
and the concordant $\Lambda$CDM model (550.354) obtained using the same set of data \cite{IHDE}.
Thus, the New HDE model does provide a nice fit to the data.

It is worthwhile to make a comparison between different models by using the information criteria (IC).
Here we adopt the Bayesian information criteria (BIC) \cite{BIC} and Akaike information criteria (AIC) \cite{AIC},
defined as
\begin{equation}
 {\rm BIC} = -2 \ln \mathcal{L}_{max}+k \ln N,\ \ {\rm AIC} = -2 \ln \mathcal{L}_{max}+2k,
\end{equation}
where $\mathcal{L}_{max}$ is the maximum likelihood satisfying $-2 \ln \mathcal{L}_{max}=\chi^2_{min}$
if assuming Gaussian errors, $k$ is the number of parameters, 
and $N$ is the number of data points used in the fit.
By comparing the NHDE and original HDE model with the original $\Lambda$CDM model with $N_{\rm eff}=3.046$,
we get the following results ,
\begin{eqnarray}
 \Delta {\rm BIC}_{\rm HDE}=5.49,&\ &\Delta {\rm AIC}_{\rm HDE}=1.11,\\
 \Delta {\rm BIC_{\rm New\ HDE}}=11.20, &\ &\Delta {\rm AIC}_{\rm New\ HDE}=2.44.
\end{eqnarray}
Notice that we define $\Delta {\rm IC}_{\rm model}\equiv {\rm IC}_{\rm model}-{\rm IC}_{\rm \Lambda CDM}$.
So, although HDE and New HDE models have slightly smaller $\chi^2$s than the $\Lambda$CDM, 
due to their extra parameters they are not favored by the ICs.

\begin{figure}
\centering{
\includegraphics[height=4.0cm]{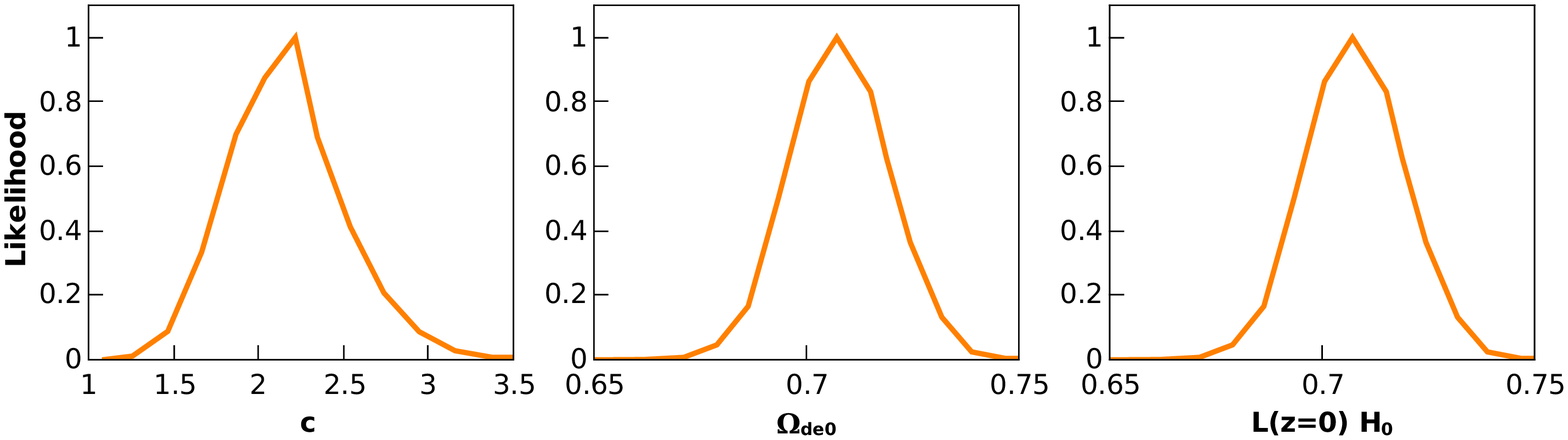}
\includegraphics[height=4.5cm]{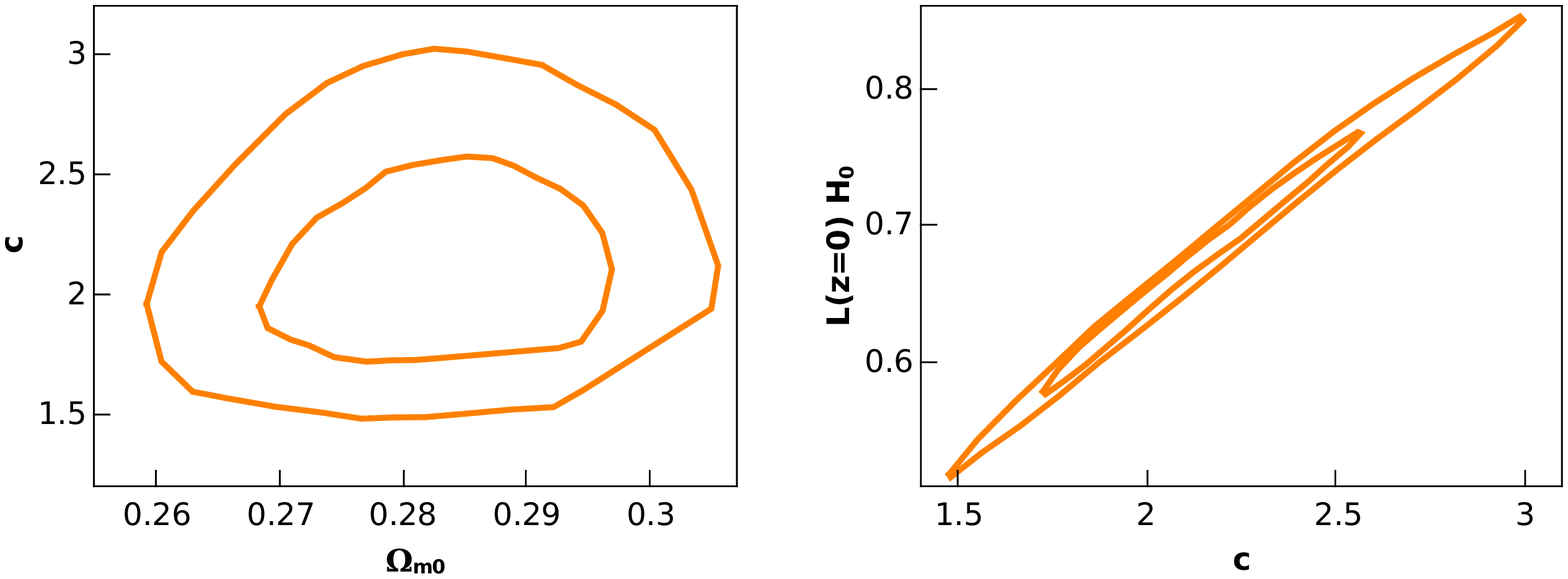}}
\caption{\label{Fig:ConLike}
Upper panels: Marginalized likelihood distribution of $c$, $\Omega_{\rm de0}$, $\tilde{L}_0$.
Lower panels: Marginalized 68.3\% and 95.4\% CL contours in the $\Omega_{\rm m0}$-$c$ and $c$-$\tilde{L}_0$ planes.}
\end{figure}

In Table \ref{table1}, we list the best-fit and errors of eight parameters,
including the five basic parameters in Eq. (\ref{Eq:ParSet})
and three derived parameters $\Omega_{\rm de0}$, $w(z=-1)$ and $N_{\rm \rm eff}$ (marked with $^*$).
The marginalized likelihood distributions of $c$, $\Omega_{\rm de0}$, $\tilde{L}_0$
are plotted in the upper panel of Fig. \ref{Fig:ConLike}.
The 68.3\% and 95.4\% contours in $\Omega_{\rm m0}$-$c$ and $c$-$\tilde{L}_0$ planes
are plotted in the lower panel of Fig. \ref{Fig:ConLike}.

\begin{table} 
\begin{center}
\label{table1}
\begin{tabular}{|c|c||c|c|}
  \hline
Parameter   &           Best fit with errors      & Parameter &     Best fit with errors   \\
 \hline
  $\Omega_{\rm m0}$  & $0.281^{+0.014+0.027}_{\rm - 0.012- 0.026}$  & $h$              & $0.724^{+0.014+0.018}_{\rm -0.023-0.040}$ \\
 \hline
  $c$          & $2.23^ {+ 0.31+ 0.85}_{\rm - 0.53- 0.82}$   & $\tilde{L}_0$& $0.70^{+ 0.07+0.18}_{\rm -0.12-0.19}$  \\
 \hline
  $\Omega_{\rm k0}$ & $0.012^{+0.003+0.010}_{\rm -0.010-0.014}$   & $w(z=-1)$   $^*$        & $-1.67^{+0.12+0.28}_{\rm -0.31-0.57}$  \\
 \hline
  $N_{\rm \rm eff}$ $^*$     & $3.54^{+0.32+0.67}_{\rm -0.45-0.76}$    & $\Omega_{\rm de0}$  $^*$  & $0.707^{+0.015+0.032}_{\rm -0.014-0.031}$ \\   
 \hline
\end{tabular}
\end{center}
\caption{Fitting results of 5 free parameters (no mark) and 3 derived parameters (marked with $^*$). }
\end{table}

We find a constraint on the parameter $c$,
\begin{equation}
1.41<c<3.09\ (95.4\% {\rm CL}),
\end{equation}
corresponding to $-2.25<w(z=-1)<-1.39$.
The $c$=0 case, 
corresponding to a Universe with matter and radiation components 
but without dark energy component,
is excluded at a high CL.
Similar to the original HDE model \cite{HDEObserv},
cosmological observations favor the Big Rip fate of the Universe in the New HDE model.
The difference is that in the New HDE model Big Rip happens at a much higher confidence level:
The upper-left panel of Fig. \ref{Fig:ConLike} shows $c<6$ in $\gg3\ \sigma$,
while in the non-flat original HDE model Big Rip only happens in 2.5$\sigma$,
as shown in Ref. \cite{HDEMCMC}.

At present, the Universe is dominated by the dark energy component with ratio
$\Omega_{\rm de0}=0.707^{+0.015+0.032}_{\rm -0.014-0.031}$
(see the upper-middle panel of Fig. \ref{Fig:ConLike} for the likelihood distribution).
The ratio of curvature is $\Omega_{\rm k0}=0.012^{+0.003+0.010}_{\rm -0.010-0.014}$,
in consistent with a flat spacetime predicted by inflation.
The current size of the future event horizon is
$L_0=0.70^{+ 0.07+0.18}_{\rm -0.12-0.19}\times H_0^{-1}$
(see the upper-right panel of Fig. \ref{Fig:ConLike} for the likelihood distribution),
slightly smaller than the Hubble radius.

The marginalized contours in the $\Omega_{\rm m0}$-$c$ plane are plotted in the lower-left panel of Fig. \ref{Fig:ConLike}.
Interestingly, we find the shape of the contour very similar to that of the original HDE model
(see Fig. 1 of \cite{IHDE} for the $\Omega_{\rm m0}$-$c$ contour for the non-flat HDE model,
plotted using exactly the same set of data),
so the role of $c$ in the New HDE model is very similar to that in the original HDE model.
The contours in the $c$-$L_0$ plane,
plotted in the lower-right panel of Fig. \ref{Fig:ConLike},
showing that these two parameters are strongly correlated to each other.

Here we briefly explain the reason for the degeneracy.
The information of the current ratio of dark radiation
is encoded in $\tilde{\lambda}_0$.
Its allowed range is small (given a set of values of $\Omega_m$, $c$ and $\Omega_{k0}$),
because its high-redshift value $\tilde{\lambda}(z_{\rm rd})$
is roughly confined to $(-10^{-4},10^{-4})$ according to Eq. (\ref{Eq:lambdazrd}).
As a result, 
the combination $c/\tilde{L_0}^2$, 
determined by $3\Omega_{de}-\frac{\tilde{\lambda}_0}{2}$ according to Eq. (\ref{Eq:inicon}),
is also constrained.
The dispersion shall be 3 times the error of $\Omega_{de}$,
which is about 0.2 at the 2$\sigma$ CL
\footnote{We can check this from the lower-right panel of Fig. \ref{Fig:ConLike}.
At $c=2$ (2.5), we get $\tilde{L}\approx0.63-0.67\ (0.73-0.75)$,
yielding $c/\tilde{L_0}^2=4.45-4.73\ (4.44-4.69)$.
The ranges are similar, and the dispersion is consistent with our estimation 0.2.
We admit that Eq. (\ref{Eq:ParSet}) is a
technically but not most physically useful parametrization of the model.
An equivalent parametrization is replacing $\tilde{L}_0$ by $N_{\rm eff}$,
which is physically more meaningful and not highly correlated with $c$ 
(see the right panel of Fig. \ref{Fig:Neff}).}.

For $N_{\rm eff}$, we get a constraint $3.54^{+0.32+0.67}_{\rm -0.45-0.76}$,
with the best-fit value slightly larger than 3.046,
corresponding to a positive dark radiation component.
We will discuss the fitting results of dark radiation in detailed in the last subsection of this section.

\begin{figure}[H]
\centering
\includegraphics[scale=0.35]{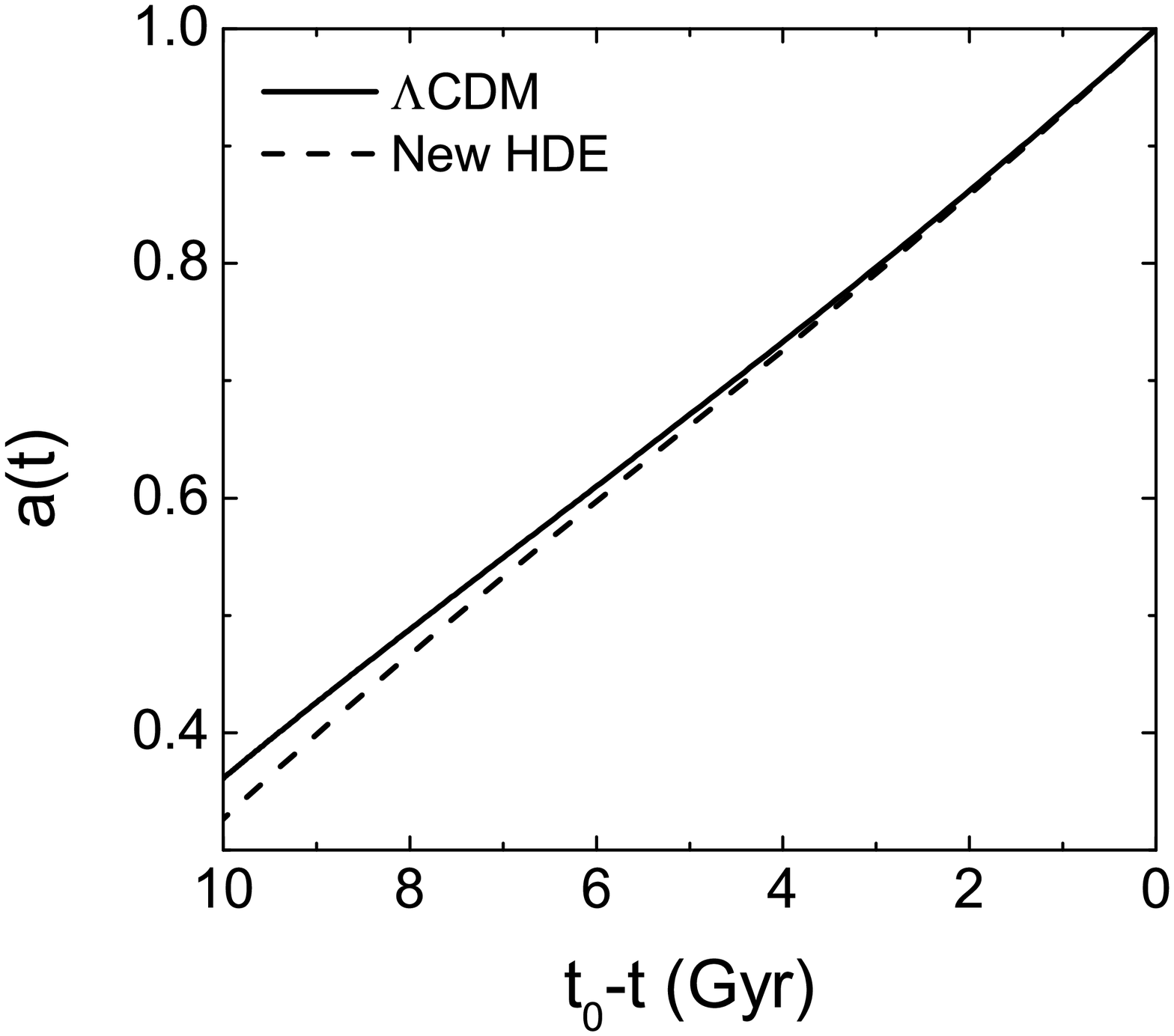}
\includegraphics[scale=0.35]{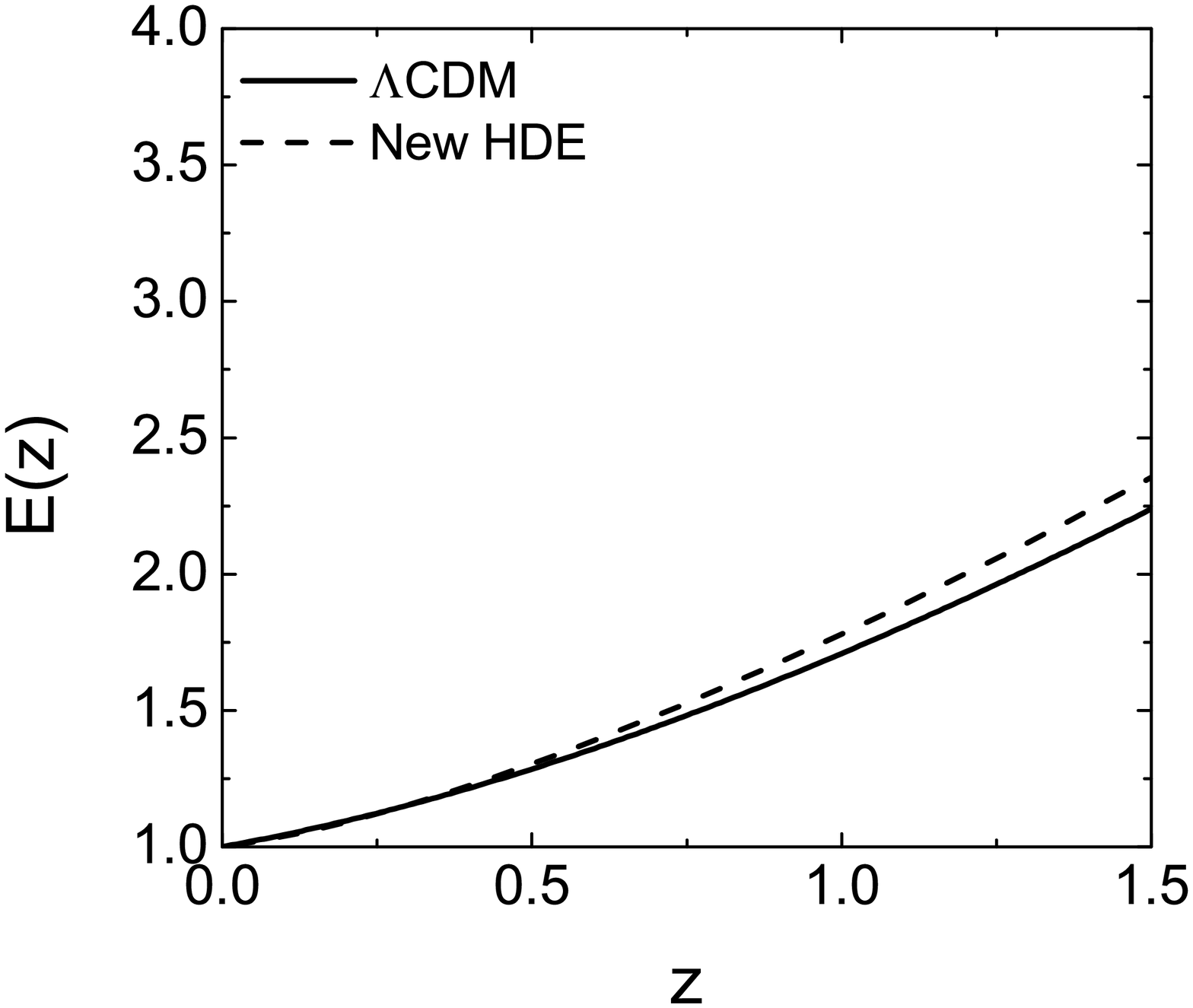}
\caption{\label{Fig:expansionhistory2}
The expansion history given by the best-fit parameters (dashed line).
The $\Lambda$CDM model with WMAP7 best-fit parameters is also plotted (solid line) for comparison.}
\end{figure}

In Fig. \ref{Fig:expansionhistory2}, 
we plot the expansion history given by the best-fit parameters (dashed line).
The WMAP7 best-fit $\Lambda$CDM model (solid line) is also plotted for comparison.
We find that, in the low redshift region, the two lines are closed to each other, 
suggesting that the New HDE model is able to provide a nice fit to the cosmic expansion history data.
In the high redshift region, the New HDE model has slightly higher expansion rate,
mainly due to the existence of a positive dark radiation component in the model.

\subsection{Equation of State}

In this subsection we discuss the EoS $w$,
which is believed to be the most important marker of the properties of dark energy.

\begin{figure}[H]
\centering{
\includegraphics[height=4.5cm]{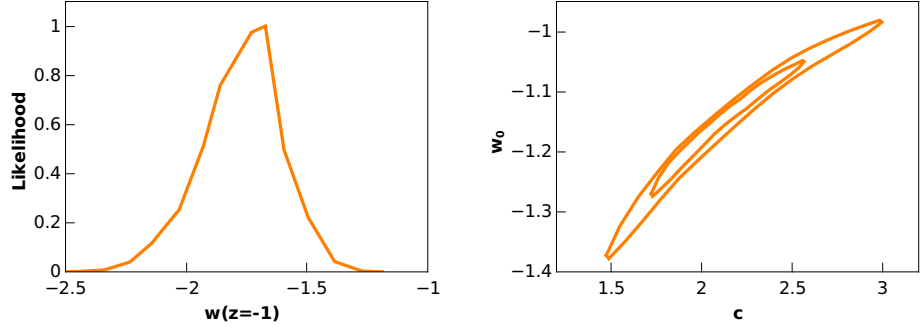}}
\caption{\label{Fig:wrelated}
Left panel: Marginalized likelihood distribution of $w(z=-1)$.
Right panel: Marginalized  68.3\% and 95.4\% CL contours in the $c$-$w_0$ plane.}
\end{figure}

As mentioned above,
the Union2.1+BAO+CMB+$H_0$ data yield the constraint $1.41<c<3.09$ and $-2.25<w(z=-1)<1.39$ (95.4\% CL),
corresponding to the Big Rip fate of the Universe.
In the left panel of Fig. \ref{Fig:wrelated} we plot the likelihood distribution of $w(z=-1)$,
which shows that $w(z=-1)<-1$ at a high confidence level.
Especially, we find the Big Rip time from now is
\begin{equation}
t_{\rm BR}=\int_{\rm -1}^0\frac{dz}{(1+z)H(z)}=20.35^{+5.01+15.87}_{\rm -7.64-9.35}{\rm  Gyr}.
\end{equation}
Thus, in the worst/most optimistic case,
our Universe can still exist for 11.0/36.2 Gyr (95.4\% CL).

For the present value of $w$,
the right panel of Fig. \ref{Fig:wrelated} shows the contours in the $c$-$w_0$ plane.
We find that $w_0\approx-1$,
with the phantom region $w<-1$ slightly favored by the data.
This figure also shows the correlation between $w_0$ and $c$,
revealing the reason {\it why $c\ll6$ from the data}:
a too large $c$ leads to a too large $w$ at $z\approx0$,
disfavored by the SNIa and BAO observations.

\begin{figure}[H]
\centering{
\includegraphics[height=7cm]{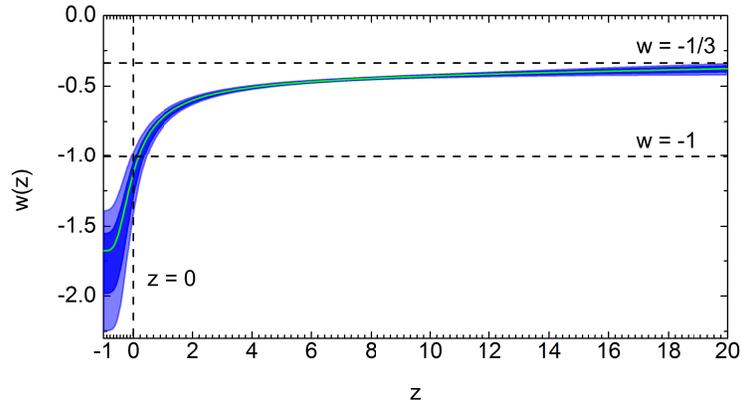}}
\caption{\label{Fig:wz}
Reconstructed evolution of $w(z)$ at 68.3\% and 95.4\% CL.
The best-fit case is plotted in green line.}
\end{figure}

For an overall view of the EoS,
in Fig. \ref{Fig:wz} we plot the reconstructed $w(z)$ at $-1\leq z \leq 20$.
We see that, $w$ cross -1 from above roughly at the current epoch,
and evolves to $w<-1$ in the future.
Moreover, in the far past we have $w\approx -\frac{1}{3}$,
a phenomenon proven in \cite{NewHDE}.
%\footnote{Another phenomena not shown in Fig. \ref{Fig:wz}
%is that we have $w\approx\frac{1}{3}$ if taking very high redshift (e.g., $z\sim1000$),
%due to the domination of dark radiation component.}.
All these properties are similar to the original HDE model.

\subsection{Dark Radiation}

In this subsection we discuss another interesting topic in the New HDE model
--- the dark radiation.
As introduced above, the dark radiation component arises naturally in the New HDE model.
and is the major new phenomenon compared with the original HDE model.

\begin{figure}[H]
\centering{
\includegraphics[height=4.5cm]{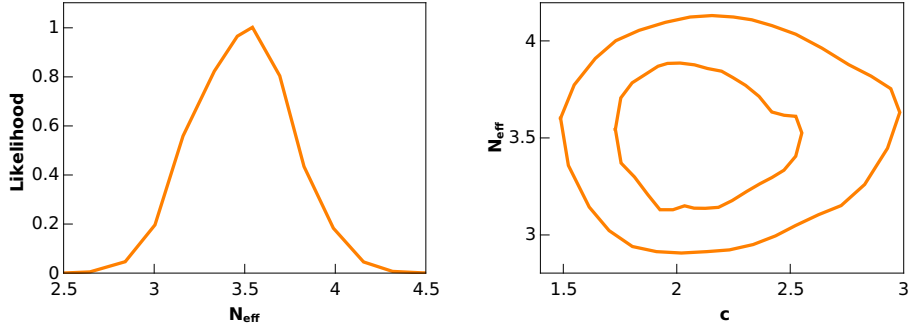}}
\caption{\label{Fig:Neff}
Left panel: Marginalized likelihood distribution of $N_{\rm \rm eff}$.
Right panel: Marginalized 68.3\% and 95.4\% CL contours in the $c$-$N_{\rm \rm eff}$ plane.}
\end{figure}

In the left panel of Fig. \ref{Fig:Neff} we plot the likelihood distribution of $N_{\rm \rm eff}$.
We find $N_{\rm \rm eff}=3.54^{+0.32+0.67}_{\rm -0.45-0.76}$,
with the central value slightly larger than the standard value 3.046.
Thus, the existence of about one specie of dark radiation is mildly favored by the data.
We find $N_{\rm \rm eff}$ basically uncorrelated with the other parameters
(see e.g. the right panel of Fig. \ref{Fig:Neff}).

\begin{figure}[H]
\centering{
\includegraphics[height=7cm]{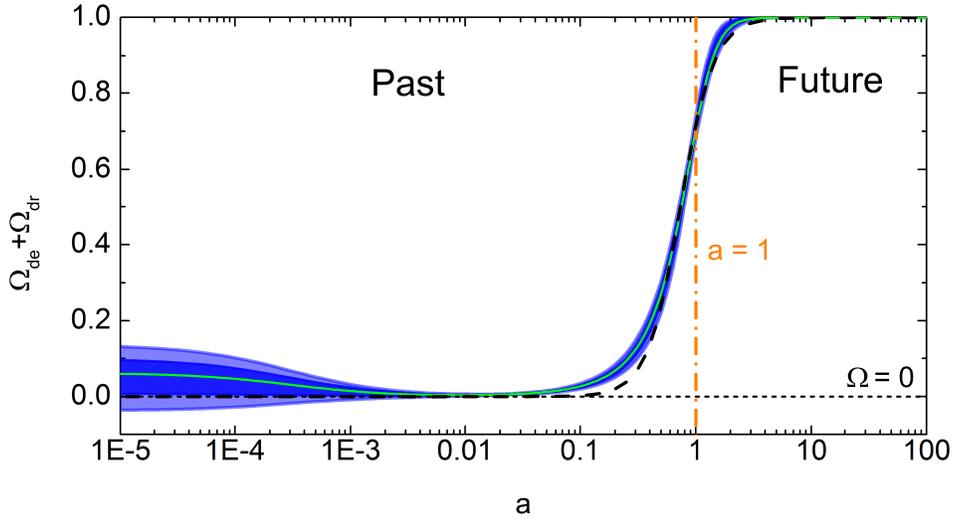}}
\caption{\label{Fig:Ode_z}
Reconstructed evolution of the ratio of the dark components (including dark energy and dark radiation) along with the scale factor $a$, at 68.3\% and 95.4\% CL,
The best-fit case is plotted in green line.
The present time $a=1$ is plotted in the thick orange line.
As a comparison, 
we also show the best-fit $\Lambda$CDM model with $N_{\rm eff}$=3.046
in thick black dashed line.}
\end{figure}

In Fig. \ref{Fig:Ode_z},
we plot the reconstructed evolution of the ratio of the New HDE along with the scale factor $a$,
from the past $a=10^{-5}$ to the future $a=100$.
We find that,
in the future we have $\Omega_{\rm hde}\rightarrow1$ due to the domination of the dark energy component,
while in the past there is an evident contribution from the dark radiation component,
with the ratio $-3.8\%\leq\Omega_{\rm dr}\leq13.5\%$ (95.4\% CL).
A negative component of dark radiation is allowed by data.

To make a comparison, we also plot the evolution 
of the dark energy ratio of the best-fit $\Lambda$CDM model 
(thick black dashed line).
We find that New HDE and $\Lambda$CDM agree with each other at the 2$\sigma$ CL in most epoch,
except the $a\approx0.05-0.5$ ($z\approx1-20$) region.
In this region, the EoS of New HDE significantly deviates from -1
(see Fig. \ref{Fig:wz}).
At the earlier epoch,
the dark energy ratio is negligible,
and the discrepancy between the two models becomes undetectable.

The dark radiation component also has evident influence on the cosmic age
$t_{\rm age}=\int_{\rm 0}^\infty\frac{dz}{(1+z)H(z)}$.
For the $\Lambda$CDM model with standard value of $N_{\rm \rm eff}=3.046$
the age is $13.76 \pm 0.11$ from the WMAP7+BAO+$H_0$ data (see \cite{WMAP7}),
while for the New HDE model we find $t_{\rm age}=13.03^{+0.56+0.93}_{\rm -0.22-0.49}$ Gyr.
Due to the dark radiation contribution,
the best-fit value shrinks, and the error bars are evidently amplified.

We find our result of $N_{\rm \rm eff}$ consistent with the previous works on dark radiation,
e.g., $N_{\rm eff}=2.79\pm0.56$ from WMAP7+ACT \cite{ACT},
and $N_{\rm eff}=3.28\pm0.40$ from WMAP9+ACT+SPT \cite{PrePlanck}.
However, it should be mentioned that our estimation of $N_{\rm \rm eff}$
obtained from the WMAP and BAO distance priors, is very rough.
To get accurate estimation on $N_{\rm \rm eff}$, one shall adopt the full CMB power spectrum data.

\section{Conclusion}

In this work we discuss the cosmological interpretations of the New HDE model \cite{NewHDE}.
Derived from the action principle,
this model overcomes the causality and circular problems in the original HDE model \cite{Li1},
and is very similar to the original HDE model,
except a new term which can be interpreted as dark radiation.

First of all, we investigate the dynamical properties and cosmic expansion history of the New HDE model.
We confirm the conclusion of \cite{NewHDE} that the equations of motion force $L(z=-1)=0$,
making $aL$ exactly the future event horizon.
We also confirm that the dark energy EoS satisfies $w\rightarrow-\frac{1}{3}$ at high redshift.
Among the six sets of model parameters considered, 
the expansion histories of three $c<6$ cases are more closed to the $\Lambda$CDM model.

Then we put constraints on the model from the Union2.1+BAO+CMB+$H_0$ data.
We get the goodness-of-fit $\chi^2_{\rm min}=548.798$,
which is comparable with the results of the original HDE model (549.461)
and the concordant $\Lambda$CDM model (550.354) obtained using the same set of data \cite{IHDE}.
Thus, the New HDE model provides a nice fit to the data,

When we assess different models by using the information criteria,
we find the New HDE model is disfavored due to its extra model parameters.
Compared with the $\Lambda$CDM model, it has a large $\Delta {\rm BIC}=11.20$.
It means that such a complicate model is not mandatory to explain the current cosmological observations.
However, theoretically it is always interesting to investigate the cosmological constraints on this model, 
and determine the region of parameter space allowed by data.

For the constraints on parameters, we get $1.41<c<3.09$ (95.4\% CL),
implying the Big Rip fate of the Universe at a high confidence level.
Correspondingly, we have $-2.25<w(z=-1)<1.39$ and the Big Rip time from now $11.0$ Gyr $<t_{\rm BR}<36.2$ Gyr (all 95.4\% CL).
For the amount of dark radiation, we get a rough estimation $N_{\rm \rm eff}=3.54^{+0.32+0.67}_{\rm -0.45-0.76}$,
with the central value slightly larger than 3.046 and a negative component of dark radiation allowed.

By reconstructing the evolution of the dark energy and dark radiation density,
we find the results of New HDE differ from the $\Lambda$CDM at $a\approx0.05-0.1$. 
In this region, the EoS of New HDE of significantly deviates from -1.
However, due to the relatively low ratio of dark energy density in the this epoch,
this difference only mildly affects the evolution of the scale factor,
as shown in Fig. \ref{Fig:expansionhistory2}.

Finally, we mention that we did not investigate the perturbations in the New HDE model.
The main objective of this paper is to investigate the region of parameter space
allowed by the cosmological observations,
and discuss the properties of this model (e.g., evolution of $w(z)$) 
based on the constraint.
The evolution of scale factor in this model
is similar to the standard $\Lambda$CDM model,
so the perturbations calculation shall not lead 
to significantly improvement on the constraint.
Moreover, the current cosmic expansion history observations are more powerful than
the growth of structure data in constraining dark energy.
Thus, in this paper we take an economical approach
and only consider the constraints from the expansion history data.

%\section{Appendix}

%\subsection{Calculation of the Big Rip time}

%The Big Rip time from now is
%\begin{equation}
%t=\int_{\rm -1}^0\frac{dz}{(1+z)H(z)}.
%\end{equation}
%At $z=-1$, $H(z)$ diverges, and we shall numerically handle this singularity.

%Let us choose
%\begin{equation}
%t=t_{\rm \Delta}+\int_{\rm -1+\epsilon}^0\frac{dz}{(1+z)H(z)}
%\end{equation}
%where
%\begin{equation}
%t_{\rm \Delta}=\int_{\rm -1}^{-1+\epsilon}\frac{dz}{(1+z)H(z)}
%\end{equation}
%and $\epsilon$ is a small number, e.g., $10^{-3}$.
%Let us focus on the calculation of $t_{\rm \Delta}$,
%which includes the singularity.

%At this redshift, we have $w\approx$constant,
%and we have (neglect all the components except dark energy)
%\begin{equation}
%H(z) \propto (1+z)^{3(1+w)/2}.
%\end{equation}
%Define $x\equiv 1+z$, we have
%\begin{equation}
%H(x) = H(z=-1+\epsilon)\times (\frac{x}{\epsilon})^{3(1+w)/2}
%\end{equation}
%and thus
%\begin{equation}
%t_{\rm \Delta}=\int_{\rm 0}^{\epsilon}\frac{dx}{x \times (H(z=-1+\epsilon) \times (x/\epsilon)^{3(1+w)/2})}
%=-\frac{2}{3 H(z=-1+\epsilon)(1+w)},
%\end{equation}
%which can be used to calculate the Big Rip time.

\begin{acknowledgments}
We are grateful to Rong-Xin Miao and Shuang Wang for useful discussions.
This research was supported by a NSFC grant No.10535060/A050207, a NSFC
grant No.10975172, and a NSFC group grant No.10821504.
\end{acknowledgments}

%%%%%%%%%%%%%%%%%%%%%%%%%%%%%%%%%%%%%%%


\begin{thebibliography}{100}

\bibitem{Riess}
A. G. Riess {\it et al.}, AJ. {\bf 116}, 1009 (1998);
S. Perlmutter {\it et al.}, ApJ. {\bf 517}, 565 (1999).

\bibitem{DEReview}
V. Sahni and A. Starobinsky, Int. J. Mod. Phys. {\bf D9}, 373 (2000);
P. J. E. Peebles and B. Ratra, Rev. Mod. Phys. {\bf 75}, 559 (2003);
T. Padmanabhan, Phys. Rept. {\bf 380}, 235 (2003);
E. J. Copeland, M. Sami and S. Tsujikawa, Int. J. Mod. Phys. D {\bf 15}, 1753 (2006);
V. Sahni and A. Starobinsky, Int. J. Mod. Phys. {\bf D15}, 2015 (2006);
J. Frieman, M. Turner and D. Huterer, Ann. Rev. Astron. Astrophys {\bf 46}, 385 (2008);
S. Tsujikawa, arXiv:1004.1493;
M. Li {\it et al.}, Commun. Theor. Phys. {\bf 56}, 525 (2011);
M. Li {\it et al.}, arXiv:1209.0922.

\bibitem{Witten:2000zk}
E.~Witten, arXiv:hep-ph/0002297.

\bibitem{Holography}
 G. 't Hooft, gr-qc/9310026; L. Susskind, J. Math. Phys. \textbf{36}, 6377
(1995); J. D. Bekenstein, Phys. Rev. D \textbf{7}, 2333 (1973); J.
D. Bekenstein, Phys. Rev. D \textbf{9}, 3292 (1974); J. D.
Bekenstein, Phys. Rev. D \textbf{23}, 287 (1981); J. D. Bekenstein,
Phys. Rev. D \textbf{49}, 1912(1994); S. W. Hawking, Commun. Math.
Phys. \textbf{43}, 199 (1975); S. W. Hawking, Phys. Rev. D
\textbf{13}, 191 (1976).

\bibitem{Cohen} A. Cohen, D. Kaplan, A. Nelson, Phys. Rev. Lett. 82,
4971 (1999).

\bibitem{Li1} M. Li, Phys. Lett. B 603, 1 (2004).

\bibitem{HDEstable}
M. Li, C. S. Lin and Y. Wang, JCAP {\bf 0805}, 023 (2008).

\bibitem{HDEObserv}
Q. G. Huang and Y. G. Gong, JCAP 0408, 006 (2004);
X. Zhang and F. Q. Wu, Phys. Rev. D 76, 023502 (2007);
M. Li, X. D. Li, S. Wang and X. Zhang, JCAP 0906, 036 (2009).

\bibitem{HDEworks}
C. J. Hogan, astro-ph/0703775; arXiv:0706.1999;
Q. G. Huang and M. Li, JCAP {\bf 0503}, 001 (2005);
X. Zhang, Int.\ J.\ Mod.\ Phys.\  D {\bf 14}, 1597 (2005); Phys.\ Lett.\  B {\bf 648}, 1 (2007); Phys.\ Rev.\  D {\bf 74}, 103505 (2006);
B. Chen, M. Li and Y. Wang, Nucl. Phys. B {\bf 774}, 256 (2007);
J. F. Zhang, X. Zhang and H. Y. Liu, Phys.\ Lett.\  B {\bf 651}, 84 (2007);
H. Wei and S. N. Zhang,  Phys. Rev. D {\bf 76}, 063003 (2007);
Y. Z. Ma and X. Zhang,  Phys.\ Lett.\  B {\bf 661}, 239 (2008);
M. Li {\it et al.}, Commun. Theor. Phys. {\bf 51}, 181 (2009);
B. Nayak and L. P. Singh,  Mod. Phys. Lett. A {\bf 24}, 1785 (2009);
K. Y. Kim, H. W. Lee and Y. S. Myung,  Mod. Phys. Lett. A {\bf 24}, 1267 (2009);
M. Li, R. X. Miao and Y. Pang, Phys. Lett. B {\bf 689}, 55 (2010); 
M. Li, R. X. Miao and Y. Pang, Opt. Express {\bf 18}, 9026 (2010);
M. Li and Y. Wang, Phys. Lett. B {\bf 687}, 243 (2010);
Y. G. Gong and T. J. Li,  Phys. Lett. B {\bf 683}, 241 (2010);
L. N. Granda, A. Oliveros and W. Cardona,  Mod.\ Phys.\ Lett.\ A {\bf 25}, 1625 (2010);
Z. P. Huang and Y. L. Wu,  arXiv:1202.4228.

\bibitem{NewHDE} M. Li, R. X. Miao, arXiv:1210.0966.

\bibitem{Kim} H. C. Kim, J. W. Lee and J. Lee, arXiv:1208.3729.

\bibitem{darkradiation}
J. Hamann et al., Phys. Rev. Lett. {\bf 105}, 181301 (2010);
J. Dunkley, et al., Astrophys. J. 739, 52 (2011);
R. Keisler, et al., Astrophys. J. 743, 28 (2011);
M. Archidiacono, E. Calabrese, and A. Melchiorri, Phys.
Rev. D 84, 123008 (2011).

\bibitem{Neffsd}
G. Mangano et al., Nucl. Phys. B 729, 221 (2005).

\bibitem{HealWorld}
X. Zhang, Phys. Lett. B {\bf 683}, 81 (2010).

\bibitem{Bassett}
C. Clarkson, M. Cortes and B. A. Bassett,
JCAP {\bf 0708}, 011 (2007).

\bibitem{Wolfram}
http://www.wolframe.com

\bibitem{Union2.1}
N.~Suzuki {\it et al.},
 arXiv:1105.3470.

\bibitem{WMAP7}
E. Komatsu {\it et al.}, ApJS. {\bf 192}, 18 (2011).

\bibitem{SDSSDR7}
W. J. Percival {\it et al.}, MNRAS {\bf 401}, 2148 (2010).

\bibitem{6dFGS}
D. H. Jones {\it et al.}, MNRAS {\bf 399}, 683 (2009);
F. Beutler, {\it et al.}, arXiv:1106.3366, MNRAS accepted.

\bibitem{AcouP}
D. J. Eisenstein {\it et al.}, ApJ {\bf 633}, 560 (2005).

\bibitem{WiggleZ}
M. Drinkwater {\it et al.}, MNRAS {\bf 401}, 1429 (2010);
C. Blake {\it et al.}, arXiv:1108.2635, MNRAS accepted.

\bibitem{HSTWFC3}
A. G. Riess {\it et al.}, ApJ. {\bf 730}, 119 (2011).

\bibitem{IHDE}
Z. H. Zhang {\it et al.}, JCAP {\bf 06}, 009 (2012).

\bibitem{BIC}
G. Schwarz, Ann. Stat. {\bf 6}, 461 (1978).

\bibitem{AIC}
H. Akaike, IEEE Trans. Automatic Control {\bf 19}, 716 (1974).

\bibitem{HDEMCMC}
Y.-H. Li {\it et al.}, JCAP {\bf 02}, 033 (2013).

\bibitem{ACT}
J. L. Sievers {\it et al.}, arXiv:1301.0824.

\bibitem{PrePlanck}
E. Calabrese {\it et al.}, Phys. Rev. D {\bf 87}, 103012 (2013).


\end{thebibliography}
\end{document}